\documentclass[aps,pre,twocolumn]{revtex4-2}
\usepackage{graphicx}
\usepackage{epstopdf, epsfig}
\usepackage{color}
\usepackage{url}
\usepackage{amsmath}
\usepackage{setspace}
\usepackage{float}
\usepackage[normalem]{ulem}

\begin{document}
\title{Nonadditive drag of tandem rods drafting in granular sediments}
\author{Brian Chang and Arshad Kudrolli}
\address{Department of Physics, Clark University, Worcester, Massachusetts 01610, USA.}

\date{\today}

\begin{abstract}
    We examine the drag experienced by a pair of vertical rods moving in tandem through a granular bed immersed in a fluid as a function of their separation distance and speed. As in Newtonian fluids, the net drag experienced by the rods initially increases with distance from the value for a single rod before plateauing to twice the value. However, the drag acting on the two rods is remarkably different, with the leading rod experiencing roughly similar drag compared to a solitary rod, while the following rod experiences far less drag. The anomalous relationship of drag and the distance between the leading and following body is observed in both dry granular beds and while immersed in viscous Newtonian fluids across the quasi-static and the rate-dependent regimes. Through refractive index matching, we visualize the sediment flow past the two rods and show that a stagnant region develops  in their reference frame between the rods for small separations. Thus, the following rod is increasingly shielded from the granular flow with decreasing separation distance, leading to a lower net drag. Care should be exercised in applying resistive force theory to multi-component objects moving in granular sediments based on our result that drag is not additive at short separation distances.
\end{abstract}

\maketitle

\section{Introduction} 
Drag and flow-mediated interactions between objects in air and water have long been studied under a variety of conditions from drafting bicyclists to bridge piling~\cite{blocken13,belloli16,stalnaker79,lachaussee18}. Their relative positions are known to have a significant effect on the drag experienced by each object in such Newtonian fluids. Flow-structure interactions are equally important in non-Newtonian fluids and granular suspensions as in mixing and blending of solids and liquids in industrial processes~\cite{coussot05}, debris flows \cite{cui2013gravity}, in bottom trawling \cite{Hiddink2017}, and biolocomotion \cite{winter14,kudrolli2019burrowing}. It is well known that for rods moving colinearly, the drag is the same as that of independent rods when they are sufficiently far apart, but the drag is reduced if they are close together at high Reynolds number flows~\cite{darabaner67a}, as well as in low Reynolds number flows when streamlines do not reconnect behind the object~\cite{moffatt63}.  At Reynolds number $Re>1$, vortices form when the cylinders are within one rod diameter as visualized in Ref.~\cite{taneda79}, and numerically at higher  $Re \sim 1-40$~\cite{vakil11}. Granular flows are different from Newtonian fluids with lower volume fractions observed behind an obstacle~\cite{buchholtz98,chehata03,albert00,brzinski10} and furrows left behind at surfaces due to the frictional rheology~\cite{cui2013gravity,allen19}. Thus, lessons learned {\color{black} on flows past cylinders in Newtonian and other non-Newtonian fluids~\cite{hu1990}} cannot be readily applied to granular flows. 

Linear superposition of elemental forces is often assumed for efficacy in calculating drag over the entire solid surface of an intruder while encountering a uniform static bed. Resistive Force Theory (RFT), originally introduced in the context of microorganism locomotion at low Reynolds number~\cite{gray55,gray09}, has been used to calculate drag acting on extended objects moving through dry granular matter~\cite{li13,zhang14,askari16}. However large systematic errors 
have been noted in comparisons with experiments and discrete element simulations \cite{li13,zhang14}. It has been suggested that the initial transient behavior when the object starts moving is the reason for the deviations. Indeed, RFT is known to lack precision in Newtonian fluids themselves, requiring empirical adjustments of the classical drag coefficients~\cite{Johnson79}. Further, the role of wake-interactions on different parts of the objects remains unclear across the quasi-static and rate-dependent regimes encountered in granular mediums. Because ageing of frictional contacts is important in granular matter, and viscous fluids drain slowly near contact points in granular beds~\cite{allen18}, the effects of a leading component may persist even when the medium has nominally come to rest.   

Here, we study the drag of two colinear rods as a function of separation distance in sediment beds. Drag in dry granular beds and also when fully submerged in water are investigated, as well as a viscous Newtonian fluid for the purpose of comparison. The drag on the two rods as a function of separation distance and speed is measured independently and compared to drag on a single rod. We demonstrate that the total drag experienced by the rods is lower in granular sediments with decreasing separation distance than in a viscous Newtonian fluid. Thus, drag acting on rods in tandem are found to be nonadditive in both dry and immersed granular sediment beds over a wide range of parameters. The leading rod is found to experience a systematically higher drag compared to the following rod at small separation distances, and even exceeding the drag of a single rod. This variation is in contrast with drag experienced in a Newtonian fluid where the drag acting on the leading rod is only slightly greater if not the same as that for the following rod at similar Reynolds numbers. To gain further insights into these results, the flow fields around the two rods are visualized with a grain-fluid refractive index-matching technique.

\section{Experimental System}
\begin{figure*}
    \begin{center}
	    \includegraphics[width=1.8\columnwidth,trim={0 11cm 2cm 0}]{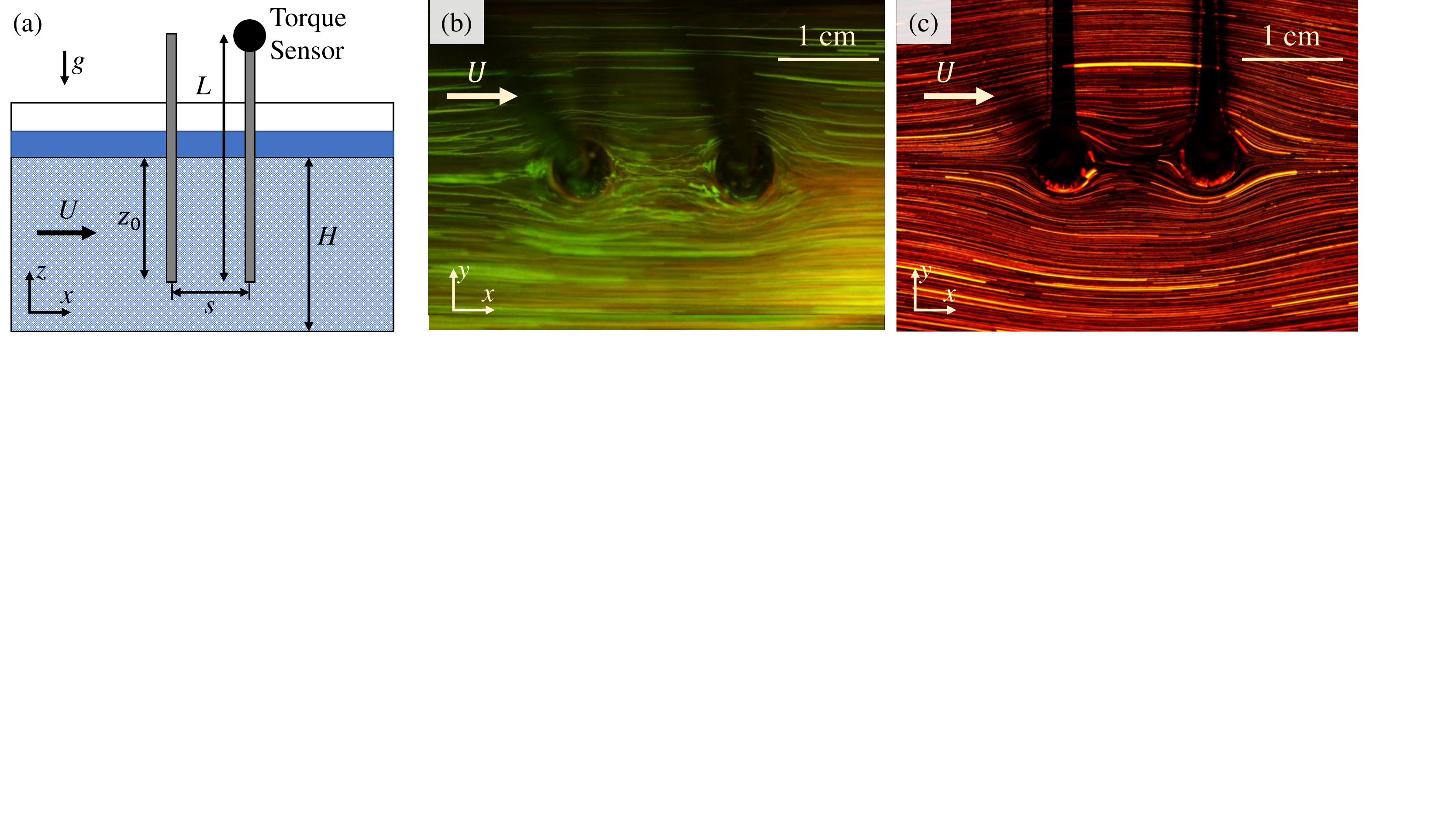}
	\end{center}
	\caption{(a) A pair of identical rods with a diameter $D$ and separated by a distance $s$ are dragged in spherical glass particles sedimented in air, water, or mineral oil. A transducer is used to measure the torque acting on one of the rods. The direction of the flow dictates whether the torque is measured on the leading rod or the following rod. 
	(b) Pathlines of borosilicate glass particles index matched with mineral oil flowing past two rods separated by $s=$1.65\,cm. 
	(c) Pathlines of flourscent tracer particles in corn syrup flowing past two rods separated by $s=$1.65\,cm with $Re=3.5\times10^{-3}$.  
	    }
	\label{fig:apparatus}
\end{figure*}
A schematic of the experimental system is shown in Fig.~\ref{fig:apparatus}(a). A pair of rods are mounted vertically inside a cylindrical container filled with granular sediments and separated by a fixed distance of $s$. The granular bed is composed of spherical borosilicate glass beads of diameter $d=1\pm0.1$\,mm with a density of $\rho_g=2.23$\,g\,cm$^{-3}$, and filled to a height of $H=4$\,cm. The rods are stainless steel with a diameter of $D=3$\,mm, hence $D = 3d$. The container has a radius of $R_c=6.1$\,cm. The rods are placed at distance of $R_t=4.0$\,cm from the container center, which is at least 10 particle diameters away from the container wall to avoid sidewall effects~\cite{allen19,allen-phd20}.  The rods are inserted to a depth of $z=3$\,cm, and at least 10 particle diameters from the container bottom.  Air, water, and mineral oil are used as interstitial fluids with physical properties listed in Table~\ref{tab:fluid}, and fill the container until the free surface is at least 5\,mm above the granular bed. This provides sufficient room above the bed such that capillary effects are absent during sediment drag experiments. In addition to experiments with the granular beds, measurements were also performed with a Newtonian fluid, light corn syrup, 
to compare and contrast with the drag of the rods.

The rods are held stationary as a rotating stage spins the container at a prescribed rate, $\omega$; their speeds in the rotating frame of reference are given by $U=\omega R_t$. The bed is initialized by rotating the rods three times around the container at a moderately fast speed, $U \sim 1.3\times10^{-1}$\,m\,s$^{-1}$, before reducing or raising to the desired speed.  In our experiments, speed is varied over three orders of magnitude from $U=1.3\times10^{-4}$\,m\,s$^{-1}$ to $U=3\times10^{-1}$\,m\,s$^{-1}$. Thus, even at the highest speeds, the centripetal acceleration $\omega^2 R_t = 0.14$\,m\,s$^{-2}$, is negligible compared to gravitational acceleration $g =9.8$\,m\,s$^{-2}$. Drag was measured after at least 10 seconds of continuous rotation to capture steady state forces. Drag is only measured on a single rod while the container rotates in a counterclockwise or clockwise direction, which gives us the drag acting on the following rod or leading rod, respectively, at various spacings. 
The total drag $F_d$ is then obtained by adding the measured drags acting on the leading and following rods. 

The Stokes Number is used to characterize the nature of particle-laden flows and is given by $St = {U \rho_g d}/{18 \eta_f}$, where $U$ is the speed of the undisturbed flow. Further, $St$ can be related to the particle Reynolds number as $St = \frac{\rho_p}{18 \rho_f} Re_p$, where $Re_p = \rho_f U d/\eta$. 
Thus, when describing suspended particles which are expected to follow the streamlines of the fluid flow, $St$ and $Re_p$ correspond to low $St$ and low $Re$. Over the parameters varied in our experiments, $St$ ranges from 1 to $2300$ in the dry bed, $0.02$ to $42$ in water, and $7.2 \times 10^{-4}$ to $1.7$ in mineral oil. $Re_p$ ranges from 0.02 to $20$ in the dry bed, $0.1$ to $300$ in water, and $4.4 \times 10^{-3}$ to $10$ in mineral oil. Thus, our investigations span regimes where viscous forces dominate and inertial effects are negligible, to where viscous forces are present and inertia starts to become important.

\begin{table}
  \begin{center}
        \def~{\hphantom{0}}
        \begin{tabular}{lccc}
        \hline
            Fluid  & $\rho_f$ (kg m$^{-3}$)  &   $\eta_f$ (mPa s) \\[3pt]
            \hline
            Air             & 1.2            &  0.018      \\
            Water           & 998             &  1          \\
            Mineral Oil     & 943             &  20         \\
            Corn Syrup     & 1400             &  2000         \\
            \hline
        \end{tabular}
        \caption{Properties of the fluids used for experiments.}
        \label{tab:fluid}
  \end{center}
\end{table}

\begin{figure*}
    \begin{center}
	    \includegraphics[width=1.5\columnwidth,trim={0cm 10cm 9cm 0}]{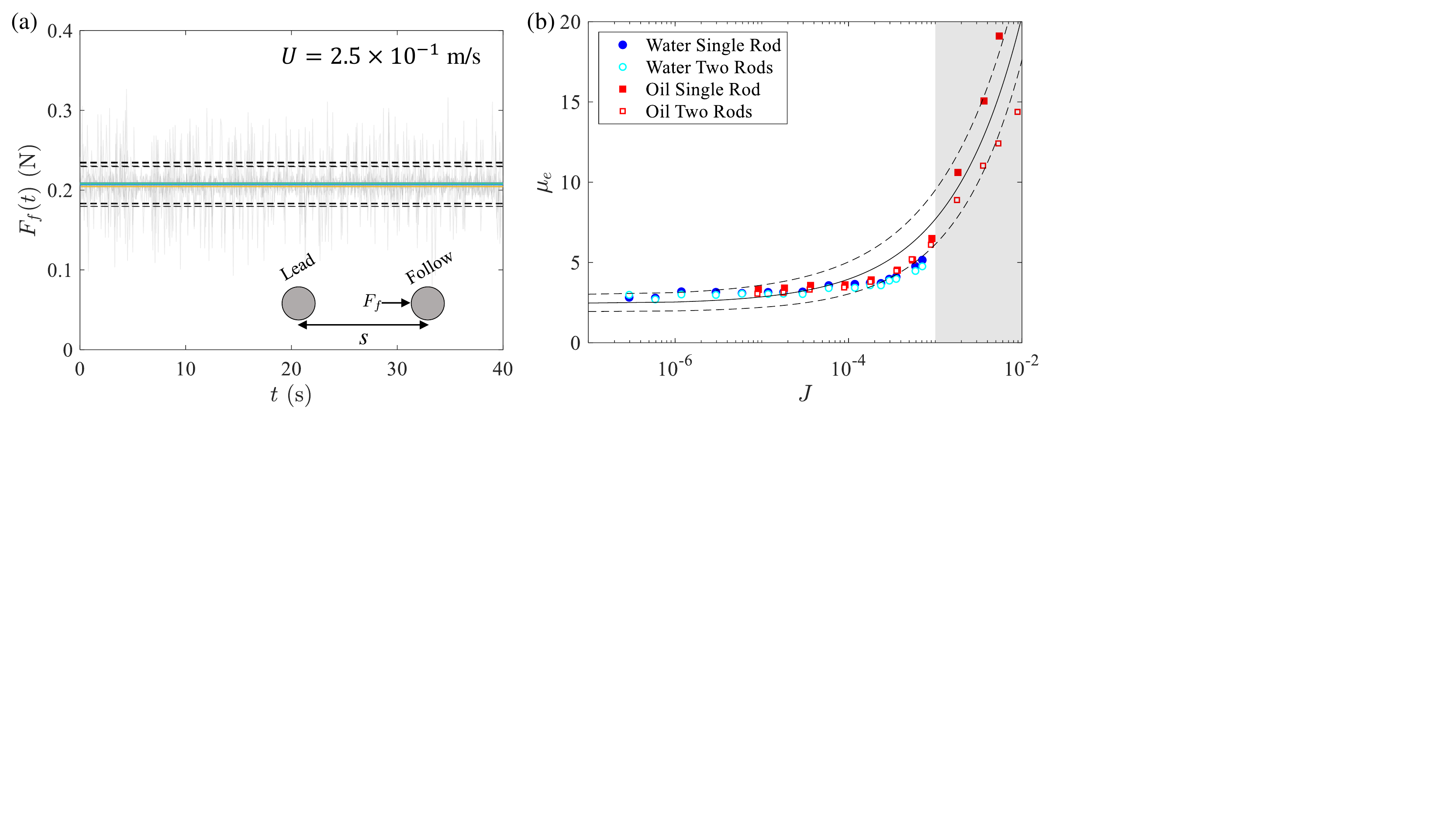}
	\end{center}
	\caption{
	(a) Time series of drag acting on the following rod moving across a granular bed immersed in water for $N=5$ trials ($s/D=40$; $U=2.5\times10^{-1}$\,m\,s$^{-1}$). The average drag and the root-mean-square-deviations are shown by solid and dashed lines, respectively. Inset: Schematic of the tandem rods and drag acting on the following rod. 
	(b) The effective friction, $\mu_e$, as a function of the viscous number $J$ for grains sedimented in the two liquids. At small speeds, $\mu_e$ approaches a non-zero value due to the yield-stress. At large speeds where $J>2\times10^{-3}$, the finite size of the container shows deviations in $\mu_e$ between single and two rods. The dashed line is fitted by $\mu_e=\mu_0+kJ^\beta$ (\cite{allen19}), where $\mu_0=2.4\pm0.5$, $k=206\mp11$, and $\beta=0.54\mp0.04$ for a single rod.
	  }
	\label{fig:force-v-time}
\end{figure*}
To gain a better understanding of the observed phenomenon and flow regime, we visualize the motion of the glass beads in refractive index-matched mineral oil and image the flow with a PL-D7512CU-T Pixelink color camera.  Rhodamine B dye, which fluoresces when illuminated with a 532\,nm light sheet, is mixed in the mineral oil.
Figure~\ref{fig:apparatus}(b) shows a sediment layer at a depth $z_o = 5d$, where the two vertical rods are also visible. The glass beads do not contain dye and thus appear dark by contrast.  However, a few of the grains have internal micro-cracks which reflect the light and act as tracers. By using a long exposure, these tracers are made visible as bright, green streaks. We observe that the granular medium moves around the rods symmetrically, with significant deviations limited to a distance of about 1\,cm, which is a few times $D$. The flow appears laminar and vortices are not apparent behind the rods. We also visualized the flow of a purely viscous fluid (corn syrup) by adding micron scale fluorescent tracers at similar $Re = 3.5 \times 10^{-3}$ in Fig.~\ref{fig:apparatus}(c). We observe that the flow appears similarly laminar, although the deviations of the flow due to the rods can be seen at further distance compared with that in the granular medium shown in Fig.~\ref{fig:apparatus}(b).

To further measure the granular flow quantitatively using Particle Image Velocimetry (PIV), we place a 532\,nm filter over the camera to remove direct reflections, and obtain a consistent image of dark grains against a bright fluid background. The images are then processed using shareware PIVlab~\cite{Thielicke_2021} to obtain the velocity field. We report those measurements after discussing the drag measurements.  

\section{Drag Measurements in Sediments}
\subsection{Single 
rod versus 
tandem rods}
We first discuss the drag acting on the two rods inside a granular bed sedimented in water. Fig. \ref{fig:force-v-time}(a) is a representative plot of the force acting on the following rod at a speed of $U=2.5\times10^{-1}$\,m/s. 
The light grey lines are the raw force data as a function of time, $F_f(t)$, for a large separation distance of $s=12$\,cm. Measurements are then averaged over approximately 40 seconds to smooth over local packing fraction fluctuations.  The solid, flat lines shown in Fig. \ref{fig:force-v-time}(a) represent the average steady state force over 40 seconds of $N=5$ trials. The dotted black lines represent the root-mean-square-error of the raw data over $N=5$ trials. Trial-to-trial variations of the mean (the number of trials, $N=5$) are less than 5\% regardless of speed, thus we perform one trial for each set of data here onward. The steady-state drags acting on the leading rod and the following rod are denoted as $F_\ell$ and $F_f$, respectively. Then, the net drag acting on the two rods $F_d=F_\ell+F_f$.

The drag acting on a single vertical rod moving horizontally across a sediment bed in air and immersed in a Newtonian liquid has been studied \cite{allen19} from the quasi-static to the rate-dependent regime. The ratio of the drag scaled by the average weight of the granular matter acting on the rod 
was found to be given by an effective coefficient of friction,  
\begin{equation}
	\mu_e(J) = \mu_0 + k J^\beta,
	\label{eqn:roddrag}
\end{equation}
where $\mu_0, k, \beta$ are material-dependent fitting constants, and $J$ is the viscous number, a dimensionless number given by the ratio of the rod speed $U$ and the Stokes settling speed $U_s$. Thus, $J = U/U_s$, with $U_s = DP/\eta$, and where 
$P$ is the mean overburden pressure due to the weight of the grains. Thus, 
\begin{equation}
    J=\eta \, U/D P.    
\end{equation}  
It may be noted that the viscous number introduced in the context of {\it steady} uniformly sheared granular suspensions~\cite{boyer11}, gives rise to the same form after assuming that the average shear rate $\dot{\gamma}$ of the flow past the rod is given by $U/D$. 

\subsection{Effect of Separation Distance}
 \begin{figure*}
    \begin{center}
	    \includegraphics[width=1.8\columnwidth, trim={0 13cm 9cm 0}]{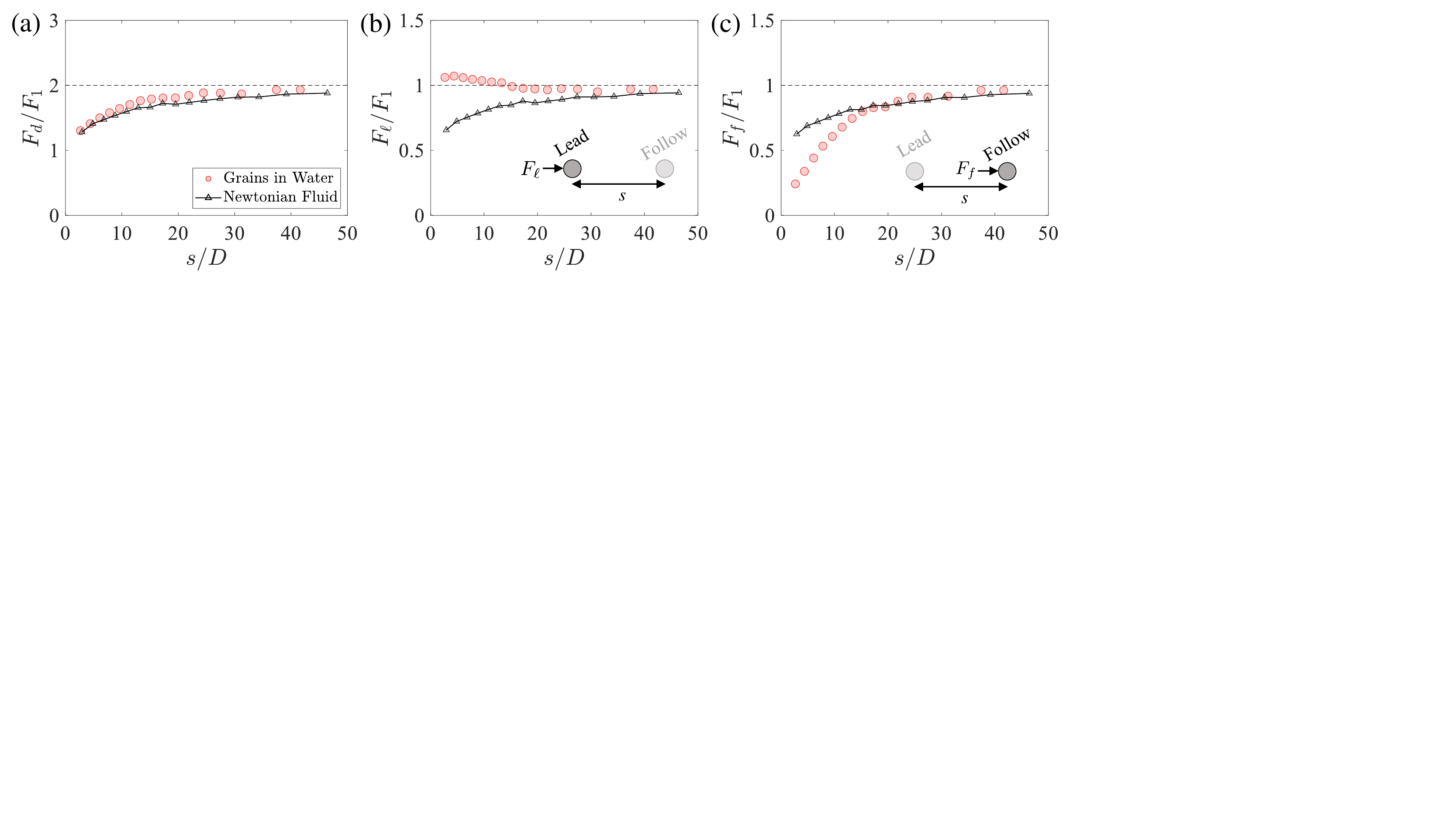}
	\end{center}
	\caption{A comparison between drag in grains sedimented in water and a viscous Newtonian fluid ($J = 8 \times 10^{-4}$). 
	(a) The net total drag acting on the two rods, $F_d$, is normalized by the total drag of a solitary rod, $F_1$. The total drag increases and approaches twice the value of a single rod with increasing separation distance, $s/D$, indicating that the forces are independent of the other rod at large separation distances. The effective Reynolds number of the sediment in water is $Re_e=8.9\times10^{-1}$ and the Reynolds number of the viscous Newtonian fluid is $Re=2.5\times10^{-1}$. (b,c) Net force acting on the following rod ($F_f$) or the leading rod ($F_\ell$) normalized by the force on a single rod $F_1$. For drag in the viscous fluid, the normalized force on the leading rod and the following rod are nearly indistinguishable with increasing separation distance. For drag in the sedimented bed, the normalized force acting on the leading rod in granular systems shows non-monotonic behaviors and is higher than the force acting on a single rod, unlike what is seen in the viscous fluid. Error bars are smaller than the size of the markers. 
	    }
	\label{fig:viscousIntruders}
\end{figure*}

We plot $\mu_e$ versus $J$ in Fig. \ref{fig:force-v-time}(b) for the rod-pair corresponding to $s/D = 40$, as well as for a single rod.  
We observe that the data collapses reasonably well on the form given by Eq.~(\ref{eqn:roddrag}), with $\mu_0=2.4\pm0.5$, $k=206\mp11$, and $\beta=0.54\mp0.04$. These material fit parameters are within the range found for a single rod by~\cite{allen19}. In the rate-independent regime, $\mu_e(J)$ is approximately the same for two rods separated at the maximum distance ($s/D \approx40$) our experiment allows. 
However, for $J>10^{-3}$, the effective friction of two rods is lower than $\mu_e$ of a single rod. We postulate that this decrease is due to the grains not fully settling before the rods come around again as they are moving on a circular path. This decrease of drag suggests a time-scale on the order of $t\sim100$\,s for the grains to come into contact. Therefore, our experiments are conducted within the rate-independent regime ($J<10^{-4}$) and the rate-dependent regime ($J=8\times10^{-4}$). Given the finite size of our experimental setup, we avoid the high $J$ regime, which yields significant differences ($ > 10$) in $\mu_e$ between a single rod and two rods separated at a large distance. 

In Fig. \ref{fig:viscousIntruders}(a), we show the net total drag acting on the two rods inside a bed sedimented in water normalized by the force acting on a single, isolated rod, $F_d/F_1$, corresponding to $J=8\times10^{-4}$. We normalize the separation distance by the diameter of the rod $s/D$. At small $s/D$, the total net drag on the two rods is 60\% less than the sum of two independent rods. This suggests there is a strong interaction between the two rods. By increasing $s/D$, the interaction between the two rods decreases as $F_d/F_1$ increases and eventually reaches a plateau toward $F_d/F_1=2$ when $s/D>30$. This shows that the drag acting on the rods are independent of each other for sufficiently large separation distances. 

To compare and contrast this observed dependence with that expected for a Newtonian fluid, we conduct the same experiment with a pure, viscous Newtonian fluid (Karo Light Corn Syrup) over a similar flow regime. The effective Reynolds number of the system is calculated as $Re_e=\rho_f U D/\eta_e$, where $\rho_f$ is intersitital fluid density, $D$ is rod diameter, $U$ is speed, and $\eta_e$ is the effective viscosity of the granular sediment. Previous work shows a relationship between $\eta_e$ and force, $F$, as follows~\cite{allen19}:
\begin{equation}
\frac{\eta_e}{\eta_f}=\frac{F}{4\pi z U}\left[ \frac{1}{2}-\mathrm{log}\left( \frac{z}{D}\right) - \mathrm{log}(4) \right].
\end{equation}
Then, the effective Reynolds number of this system is $Re_e \approx 0.9 $, 
 and from $St = \frac{\rho_p}{18\rho_f} Re_p$, we have $St \approx 0.12$, i.e. in the low Stokes number regime. Thus, 
 we perform similar measurements in the Newtonian fluid at a similar Reynolds number of $Re=0.25$. The normalized total net drag as a function of separation, also plotted in Fig.~\ref{fig:viscousIntruders}(a), shows that the drag in grains sedimented in water is actually quite close to the pure Newtonian fluid. This may lead one to believe that matching $Re_e$ of the sediment with $Re$ of a pure viscous fluid would have the same flow physics. However, decomposing and examining the forces on the leading and following rods individually, as shown in Fig.~\ref{fig:viscousIntruders}(b) and (c), respectively, reveals very different behavior, indicating that cooperative effects of two rods in tandem in a sediment is vastly different than in a viscous fluid. 

\begin{figure}
    \begin{center}
	    \includegraphics[width=.7\columnwidth, trim={0 8cm 19cm 0}]{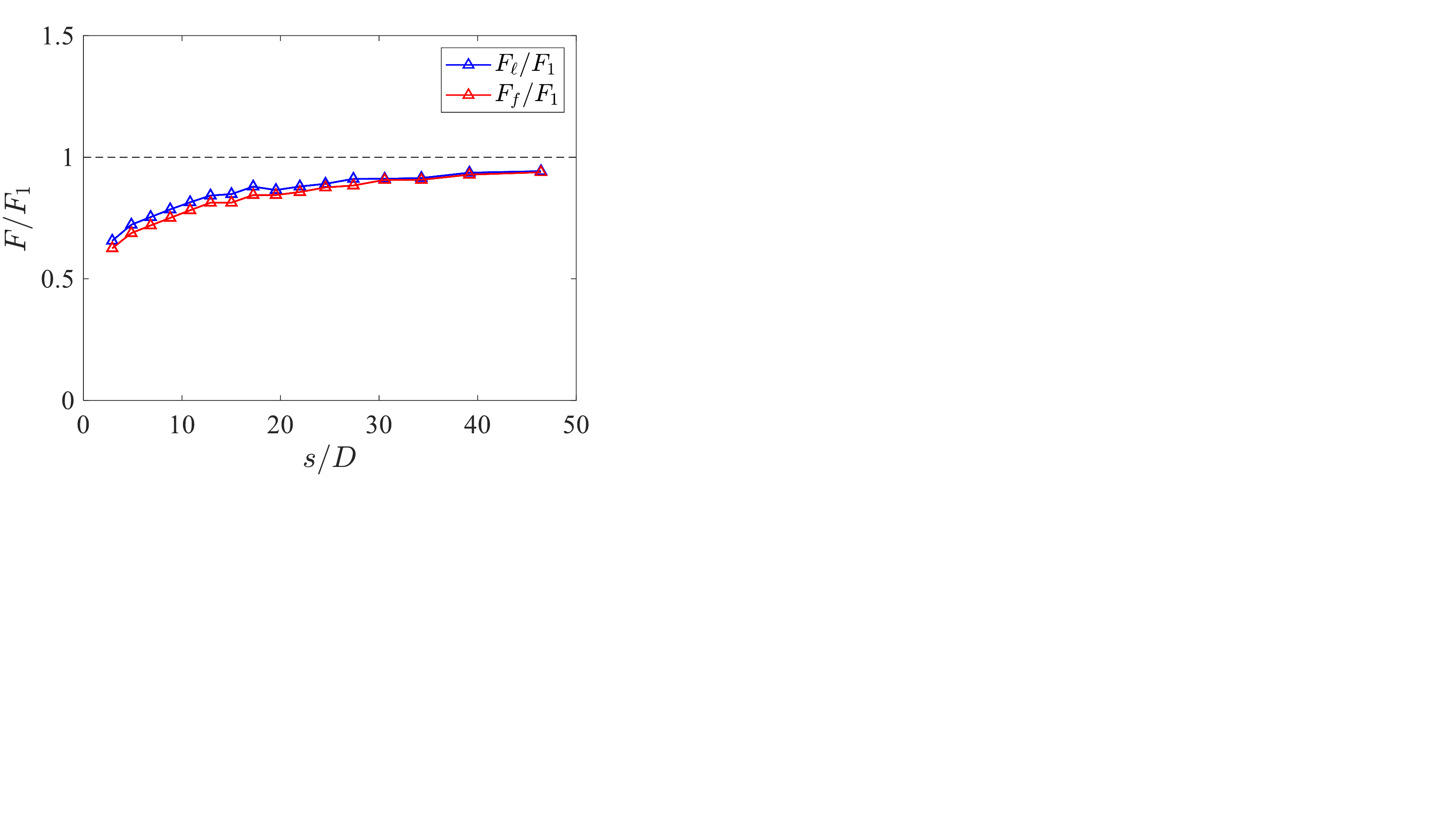}
	\end{center}
	\caption{Force acting on the leading rod $F_\ell$ and the following rod $F_f$ normalized by the force acting on a single rod $F_1$ in a viscous Newtonian fluid. $F_\ell$ tends to be slightly higher than $F_f$ at lower $s/D$ but are overall nearly indistinguishable from one another ($Re = 2.5 \times 10^{-1}$).
	    }
	\label{app:viscousforce}
\end{figure}

In the Newtonian fluid, the normalized forces on the following rod $F_\ell/F_1$ and the leading rod $F_f/F_1$ are nearly identical with increasing separation distance, as shown further in the direct comparison plotted in Fig.~\ref{app:viscousforce}. This near symmetry between the leading rod and following rod forces is due to the near fore-aft symmetry of the flow around the cylinder at low $Re$~\cite{tritton59}.  

In the sediments, $F_f$ is 37\% less at small separation distances compared to the force on an independent rod, and the drag on the leading rod is systematically greater, and even up to 6\% greater compared to the drag acting on an independent rod at relatively small separation distances ($s/D<10$). 
I.e., there is a non-monotonic behavior, such that a peak in $F_\ell$ occurs at $s/D=4.4$ before approaching unity at $s/D=10$.  

Further, $F_f$ is systematically lower (up to 60\% lower) than the drag acting on an independent rod at $s/D<30$. 
While this may seem more similar to drafting in a fluid, the force decrements in the granular sediment are much larger at small distances than the force decrements in the Newtonian fluid. $F_f/F_1$ then approaches unity with increasing $s/D$ for both cases. However, we note that the plateau occurs much later $s/D>30$ for the following rod than for the leading rod $s/D>10$ in the granular sediment case. 

We also observe that the force on the following rod becomes independent of the leading rod much faster in the sediment than in a viscous fluid. The leading rod exhibits the same peak around $s/D=4.4$ in the rate-independent regime. But in the rate-dependent regime, we find that $F_\ell/F_1$ approaches unity at small $s/D$. 
The large discrepancy between the forces in the leading rod and the following rod leads us to believe that time-irreversible effects due to inter-particle friction is responsible at low $s/D$. In the next section, we will see how various speeds and interstitial fluids in the sediment may affect the force behaviors. 

\subsection{Effect of Speed and Interstitial Fluid}
\begin{figure*}
    \begin{center}
        \includegraphics[width=1.8\columnwidth, trim={0cm 2cm 5cm 0}]{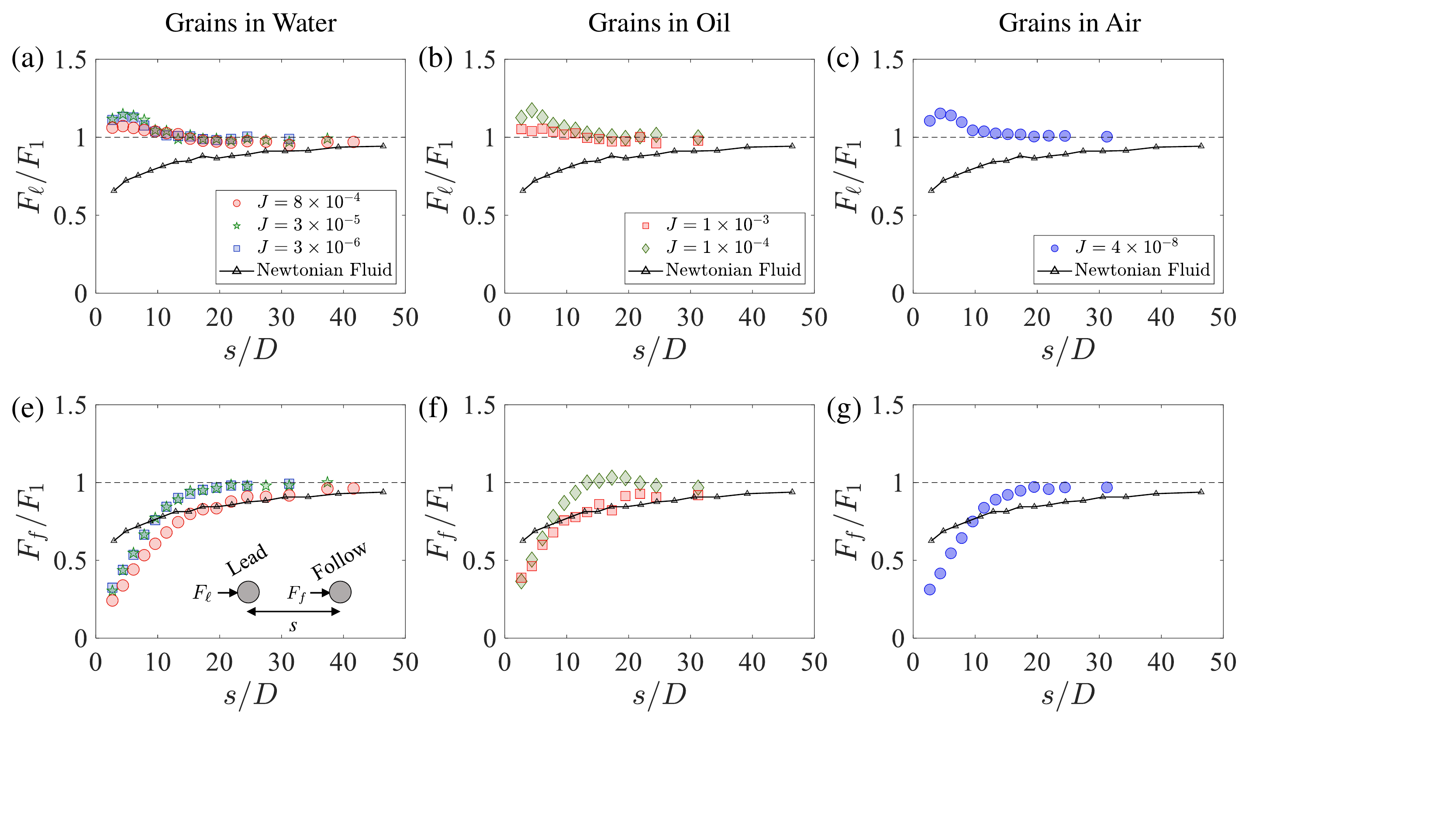}
    \end{center}	
	\caption{(a-c) Net force acting on the leading rod, $F_\ell$, normalized by the force on a single rod, $F_1$, for various $J$ and interstitial fluids. The non-monotonic behavior of the leading rod is prevalent for low $s/D$ regardless of the interstitial fluid in the rate-independent regime. This behavior becomes less pronounced at small $s/D$ as $J$ increases toward the rate-dependent regime. This is contrary to what is observed in a viscous Newtonian fluid. (e-g) Net force acting on the following rod, $F_f$, normalized by the force on a single rod, $F_1$. The force on the following rod increases monotonically with $s/D$ until reaching a plateau, at which point the following rod acts independently from the leading rod. Approaching the rate-dependent regime, the plateau begins at larger $s/D$.
	    }
	\label{fig:all}
\end{figure*}

We examine the robustness of these phenomena by changing the speed and the interstitial fluid of the sediments. The two rod system is decomposed into the leading rod and the following rod, as shown in Fig. \ref{fig:all}. For comparison, the drag corresponding to the low-$Re$ Newtonian fluid acting on the leading and following rods are also shown.

For grains sedimented in water, we decrease the viscous number $J$ toward the rate-independent regime and compare the forces acting on the leading and following rod in Fig.~\ref{fig:all}(a),(e). As previously discussed, $F_\ell$ in the rate-dependent regime is slightly higher than $F_1$ at low $s/D$. We decrease $J$ by decreasing the speed of the rods and show that within the rate-indepedent regime, $F_\ell$ is nearly 15\% higher than $F_1$ (see Fig. \ref{fig:all}(a)). In fact, the behavior of $F_\ell/F_1$ with increasing $s/D$ is nearly identical at sufficiently low $J$. Moreover, $F_\ell/F_1$ has a non-monotonic response with for $s/D<10$, and then plateaus toward unity. The non-monotonicity appears robust regardless of the interstitial fluid, as long as the viscous number is sufficiently low such that $J<1\times10^{-3}$ (see Fig. \ref{fig:all}(a-c)). For example, when the grains are submerged in oil, the force on the leading rod $F_\ell/F_1$ exhibits a non-monotonic response at $J=1\times10^{-4}$ at low $s/D$ before reaching a plateau for $s/D>10$ (see Fig. \ref{fig:all}(b)). Increasing $J$ suppresses the non-monotonic response. For dry granular media, as seen in Fig. \ref{fig:all}(c), this response is also prominent. However, we are unable to increase $J$ into the rate-dependent regime for the dry granular case due to limitations of our setup. But it is representative of the lower limit of the rate-independent regime.

For drag on the following rod $F_f/F_1$, with the grains in water, we find that the shielding effects are dependent on $J$. In the rate-dependent regime, $F_f/F_1$ is systematically lower than what is observed in the rate-independent regime (see Fig.~\ref{fig:all}(e)). Additionally, the forces in the rate-independent regime approach unity at approximately $s/D=20$ when the grains are submerged in water or air. We postulate that at low $J$, the grains have more time to fully settle before interacting with the following rod, which would lead to higher forces due to frictional effects of settled grains. A similar effect is seen when the interstitial fluid is oil as shown in Fig.~\ref{fig:all}(f). 

Comparing these results with the Newtonian fluid at low-$Re$ shows clear differences. The behavior of forces on the leading rod, regardless of speed or interstitial fluid, in a granular sediment is dramatically different from a low-$Re$ Newtonian fluid at low separation distances. At high enough $s/D$, both granular sediments and Newtonian fluid cases eventually converge to unity. The forces on the following rod in the Newtonian fluid increases with separation distance at a lower rate than the granular sediment case. 

Since interstitial fluid does not affect the overall behaviors, this leads us to believe that neither lubricating effects nor pore pressure is driving this behavior asymmetry in forces on the leading and following rod. The viscous number $J$ incorporates the amount of overburden-pressure which affects the amount of frictional contact. Therefore, we propose that inter-particle friction is the primary driver for the phenomenon observed in two drafting rods. We will further discuss this by examining the flow fields.

\section{Flow visualization}

\begin{figure}
    \begin{center}
	    \includegraphics[width=0.75\columnwidth,trim={0 0cm 24cm 0},clip]{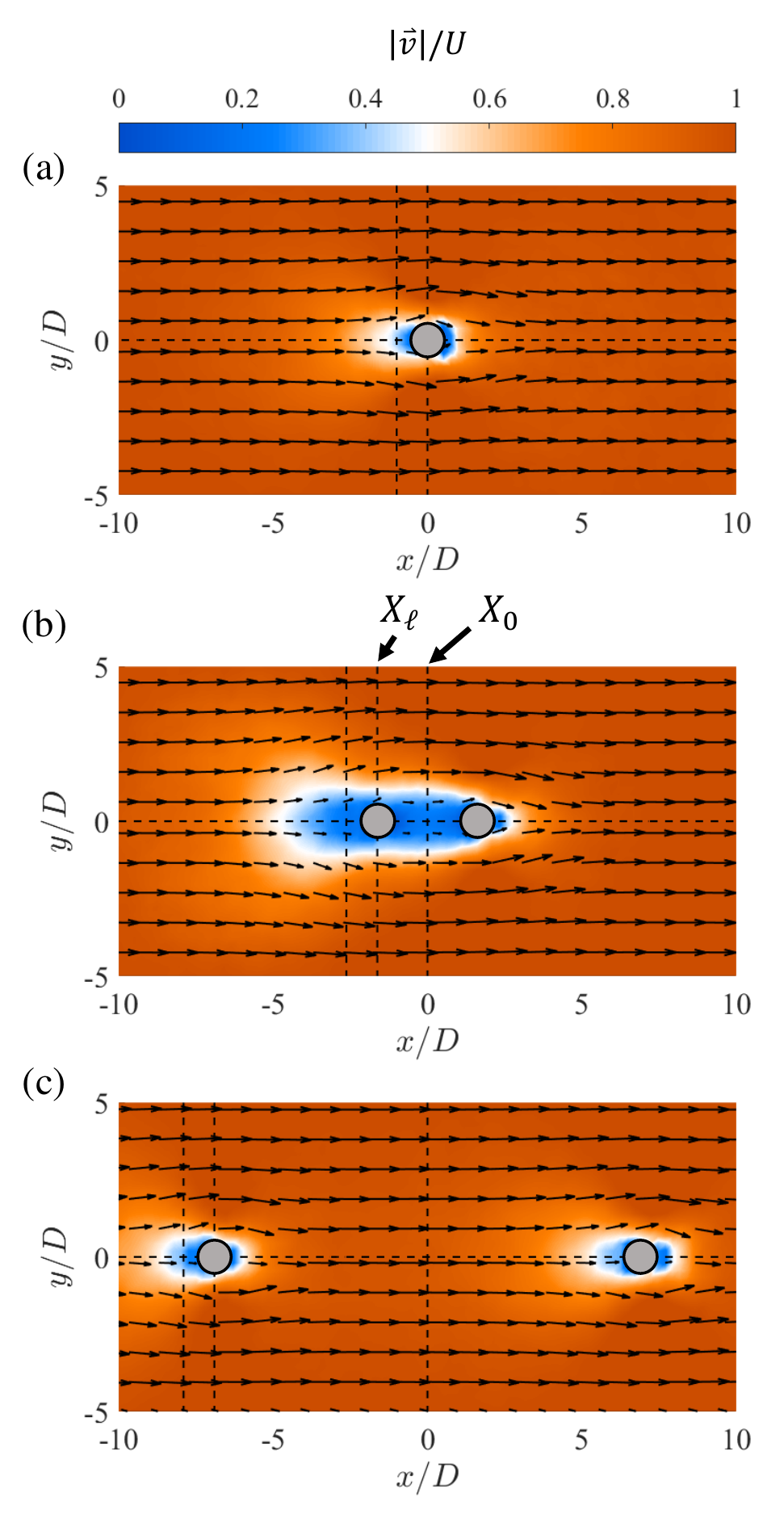}
	\end{center}
	\caption{Mean flow field normalized by the imposed velocity of $U=1.3$\,mm/s or about $0.4D/$s at depth $z_o = 5d$. Flow is moving from left to right, as indicated by the velocity vectors. (a) Flow past one rod shows a left-right-asymmetry, where particles slow down in a conical shape in front of the rod. (b) Flow past two rods with $s/D=5.5$. A strong interaction between the two rods is noted by the region between the rods that is 50\% slower than the imposed velocity. (c) Flow past two rods with $s/D=11$. The flow field around the two rods are independent from each other. The two vertical dotted lines $X_0$ and $X_1$ are denoted as the line equidistant from the two rods and the line that passes through the leading rod, respectively. The horizontal dotted line passes through the two rods along $y/D=0$. Velocity profiles along these lines are shown in Fig. \ref{fig:ux-y} and \ref{fig:ux-x}. }
	\label{fig:PIV}
\end{figure}

\subsection{Mean Flow Fields}
The mean flow field of the granular media moving around one and two intruders is shown in Fig. \ref{fig:PIV}. The mean flow field is obtained using PIVlab~\cite{thielicke2014pivlab} with an interrogation window of $6d \times 6d$ with a 50\% step per pass and three passes. The refractive index matching allows us to view a depth of $z_o=5d$. The flow fields are reflected about the y-axis due to stronger image quality near the laser to show top-bottom symmetry, which is expected in a granular flow around an obstacle with zero pressure gradient~\cite{guillard2014lift}. Velocity magnitude is normalized by the imposed velocity of $U=1.3$\,mm/s. We first analyze the velocity field for a single rod moving through grains sedimented in oil. Given the constraints of the linear stage motor, we are only able to conduct linear visualization experiments within the rate-independent regime at $J<10^{-4}$. As shown in Fig. \ref{fig:PIV}(a), a large region of low velocity ($|\vec{v}|/U<0.5$) flow develops in front of the rod in a conical shape, whereas a smaller low velocity region develops behind the rod. This is similar to what one might observe for a cylinder moving through a dense granular medium in other configurations~\cite{chehata03,guillard2014lift}. When we introduce a second rod into the system in Fig.~\ref{fig:PIV}(b) with a separation of $s/D=5.5$, the region fore of the leading rod and aft of the following rod exhibits a similar respective flow field as a single rod. However, the region in between the two rods show a semi-stagnant zone such that the velocity is less than half of the imposed velocity $U$. This indicates that the two rods are interact, and thus altering their drags. Increasing the separation distance further to $s/D=11$, we note that the flow fields around the leading and following rods are similar to that of a single rod, indicating that the rods are acting independent of one another consistent with the measured drag. 

We examine the velocity of the sediments between the rods, along and across the flow, to understand the effect of the rods as a function of separation distance more critically.

\subsection{Velocity Profile Along $x/D=0$}
\begin{figure*}
    \begin{center}
	    \includegraphics[width=1.5\columnwidth,trim={0 0cm 17cm 0},clip]{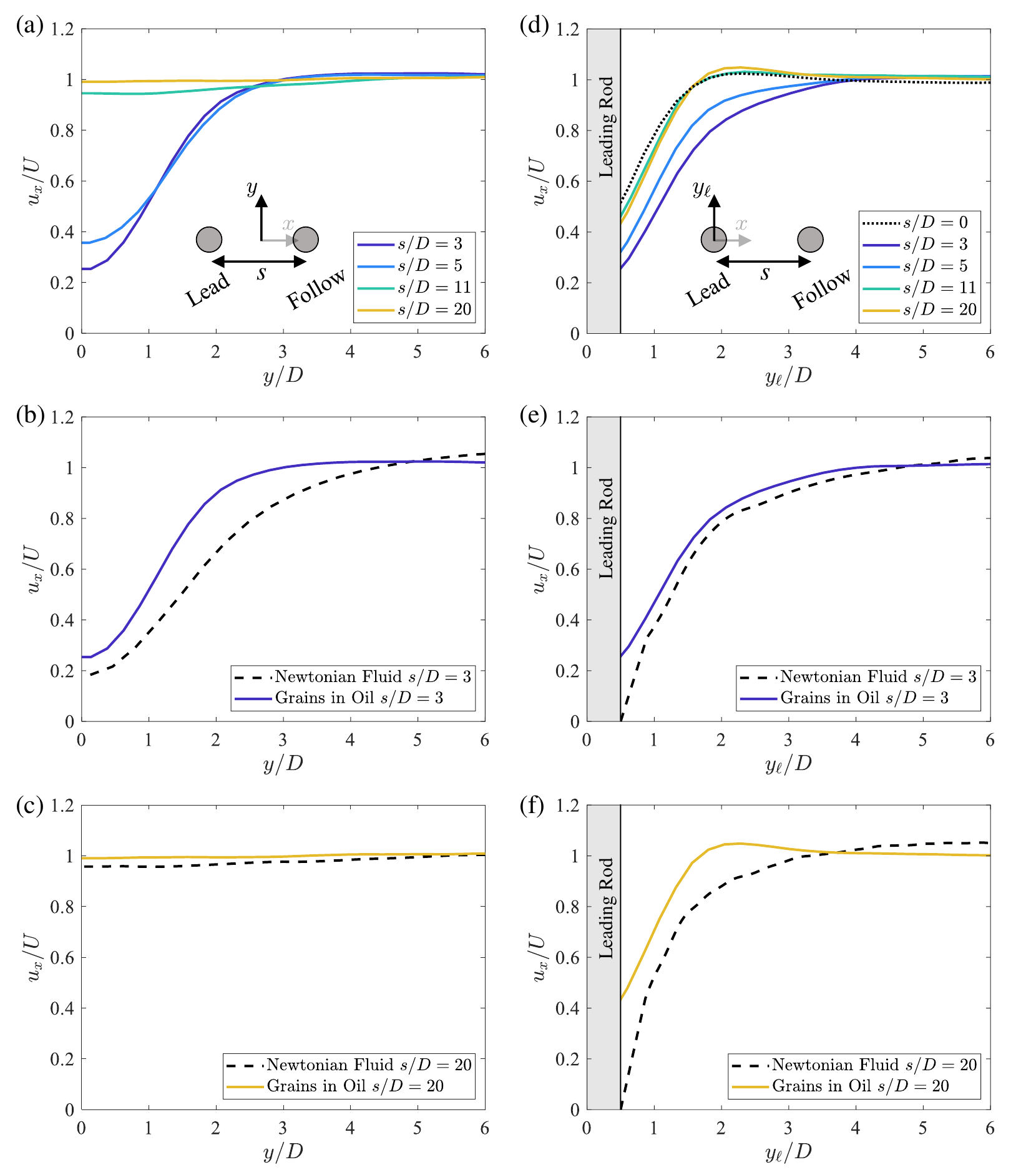}
	\end{center}
	\caption{(a) Velocity profile, $u_x/U$, along the dotted line $X_0$ (midpoint between the two rods, see Fig.~\ref{fig:PIV}) for various separation distances. Increasing separation between the rods induces a transition from a nonuniform to a uniform velocity profile. (b) Comparison between a Newtonian fluid and granular sediment at $s/D=3$ along $X_0$. (c) Comparison between the Newtonian fluid and granular sediment at $s/D=18$ along $X_0$. (d) Velocity profile, $u_x/U$, along the dotted line $X_1$ (vertical line through the leading rod) for various separation distances. 
	(e) Comparison between a Newtonian fluid and granular sediment at $s/D=3$ along $X_1$. (f) Comparison between a Newtonian fluid and granular sediment at $s/D=20$ along $X_1$. The comparisons in (e,f) required a linear interpolation toward a zero velocity due to the non-slip boundary condition.
	    }
	\label{fig:ux-y}
\end{figure*}

For the two rod system, we demarcate a line between the two rods at $x/D=0$ (shown as a black dotted line in Fig.~\ref{fig:PIV}(b),(c)), and spatially average the velocity in the x-direction over $x/D=3$ to obtain the velocity profile along the y-axis, as shown in Fig. \ref{fig:ux-y}(a). For a small separation of $s/D=3$, the velocity drops to $0.1U$ at the midline and increases towards $U$ as we move away from the midline. When $s/D=5.5$, the interaction between the two rods decreases, but is still prevalent as the velocity decreases nearly 60\% from $U$. With increasing separation distance, we note that the velocity at the midline increases towards $U$. In general, $u_x$ seems to reach the imposed velocity around $y/D=3$. This indicates the length scale at which the wake from the leading rod interacts with the following rod.
At a far enough separation such as $s/D=20$, the velocity profile hovers near $u_x/U=1$ indicating low interaction between the two rods. Such behavior is consistent with our observations of drag on the rods in an oil sediment, such that $F_\ell/F_1$ and $F_f/F_1$ are nearly 1 when $s/D=20$.

In Fig.~\ref{fig:ux-y}(b) and (c), we compare velocity profile $u_x/U$ of the Newtonian fluid with the sedimented bed for separation distances of $s/D=3$ and $s/D=20$, respectively. At $s/D=3$, $u_x/U$ is nearly $0.2$ at the midline and increases toward unity at a slower rate than the granular sediment. \textcolor{black}{This suggests that the granular sediment must be dissipating energy faster than the Newtonian fluid.} At $s/D=20$, $u_x/U$ for the granular sediment has plateaued near unity, whereas $u_x/U$ for the Newtonian fluid still exhibits a slope. This is again consistent with what we observe in the forces in a low-$Re$ Newtonian flow where a larger separation is required to reach two rod independence than in a granular sediment. This further supports the idea that inter-particle friction allows a faster rate of energy dissipation than in the low-$Re$ Newtonian fluid. 

\subsection{Velocity Profile Along $x/D=X_\ell$}
In Fig.~\ref{fig:ux-y}(d), we plot the velocity profile across the leading rod, which is demarcated by the line $X_\ell$ in Fig.~\ref{fig:PIV}(c) and (d). Here, we can compare how the velocity profile across the leading rod of a two rod system deviates from a single rod. Operating in a granular sediment, we acknowledge that surface slip effects exist. Starting with the single rod case, or $s/D=0$, we note that the velocity profile near the leading edge surface starts near $u_x/U=0.5$, overshoots unity around $y_\ell/D=2$, then decreases and plateaus to unity at $y_\ell/D=4$. Such behavior is consistent with \cite{guillard2014lift}. However, the introduction of the second rod at a separation of $s/D=3$ or $s/D=5$ decreases the initial velocity at the leading rod surface and eliminates the overshoot. The velocity profile increases until reaching unity at $y_\ell/D=4$. This indicates that the following rod has interrupted the leading rod's velocity profile.
When exploring the velocity profile with high separation distances such as $s/D=11$ and $s/D=20$, we find that the velocity profile nearly matches that of a single rod. This indicates that the introduction of a second rod has minimal effect on the leading rod only at high separation distances. This is consistent with what we observe in the force data.

It is noteworthy that the velocity at the surface of the leading rod at small separations is lower than that of the leading rod and high separations. The influence of the following rod seems to introduce a region of semi-stagnant flow around the leading rod, as shown in Fig.~\ref{fig:PIV}(b). This is a possible mechanism behind the non-monotonic behavior observed in the Fig.~\ref{fig:all}(b) for low viscous numbers. \textcolor{black}{However, it is unclear if inter-particle friction is the primary cause, or if there is a more complex interplay with particle size and depth of the rods.}

In Fig. \ref{fig:ux-y}(e) and (f), we compare the velocity profile $u_x/U$ of the Newtonian fluid with the sedimented bed for separation distances of $s/D=3$ and $s/D=20$, respectively. The primary difference noted here is that the viscous Newtonian fluid must experience the no-slip boundary condition. Therefore, the velocity is driven to $0$ at the surface of the rod and linearly interpolated to the measured velocity one-$D$ from the surface. At $s/D=3$, there is a small difference between the Newtonian fluid and the grains in oil other than due to the surface boundary condition. At $s/D=20$, the profile of the Newtonian fluid increases until it plateaus towards unity at $y_\ell/D=4$, whereas the grains in oil case exhibits an overshoot before the plateau. This is not reflected in the leading rod force data, $F_\ell/F_1$, in Fig. \ref{fig:all}(b), where the difference is greatest at $s/D=3$, and the smallest difference is at $s/D=20$.

\subsection{Velocity Profile Along $y/D=0$}

\begin{figure*}
    \begin{center}
	    \includegraphics[width=1.5\columnwidth,trim={0 8cm 5cm 0}]{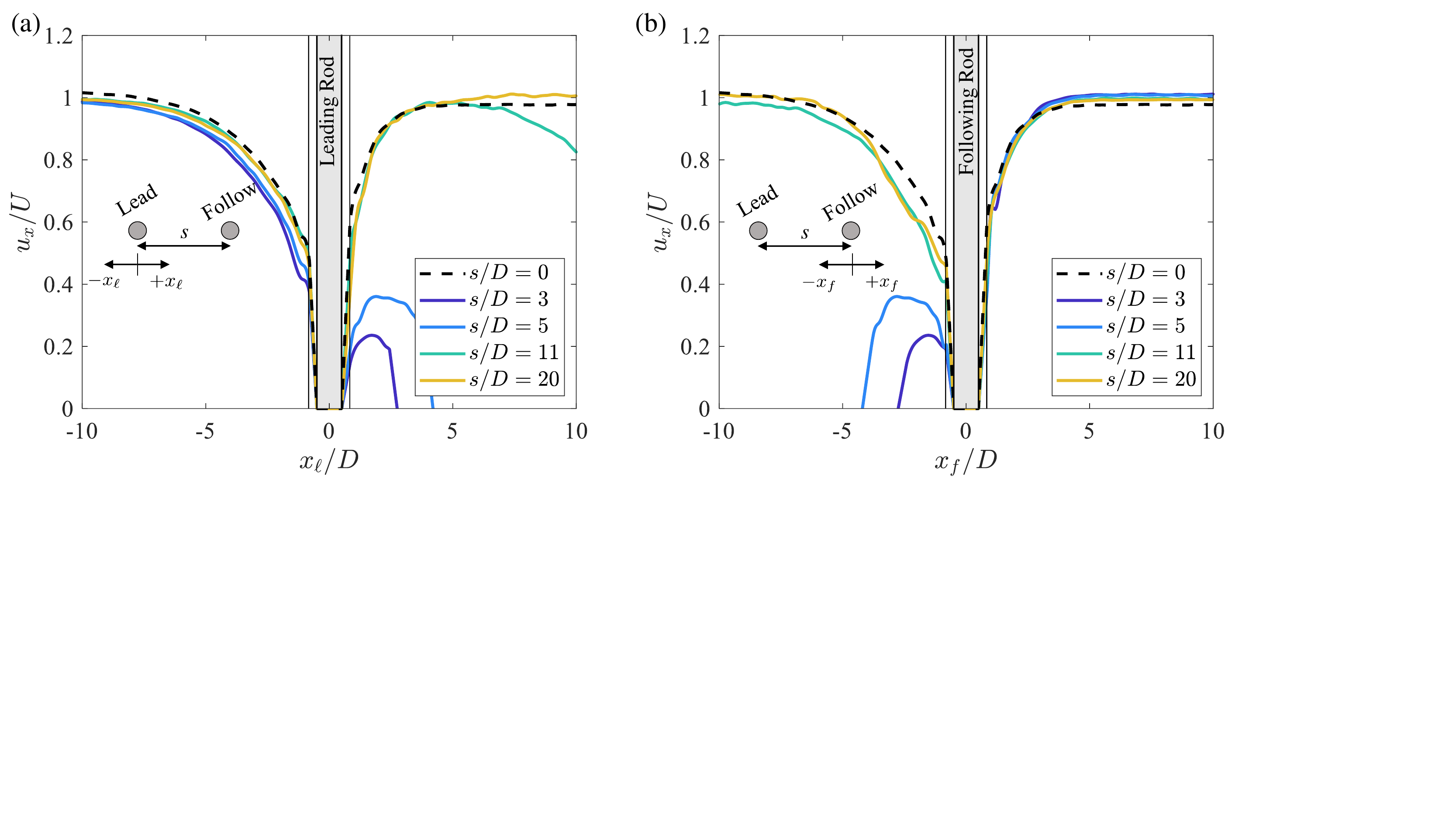}
	\end{center}
	\caption{Velocity profiles of $u_x/U$ along the midline $y/D=0$ with varying rod separation distances, $s$, from the (a) leading rod and (b) following rod frame of reference. For comparison, the velocity profile of a single rod is shown and denoted as $s/D=0$. The fore and aft profiles are noted by $-x$ and $+x$, respectively. Solid vertical lines near the rod boundaries represent the one particle diameter region from the fore and aft faces of the rods, in which the velocity profiles are forced toward zero due to a hard-repulsion boundary condition.
	    }
	\label{fig:ux-x}
\end{figure*}
Next, we investigate the velocity profile along the midline $y/D=0$ in Fig.~\ref{fig:ux-x}. The velocity profile $u_x/U$ is divided into two sections, each respective of the flow relative to fore and aft of the leading rod and the following rod. To satisfy the hard surface boundary condition, grains directly in front of and behind the rods should not have a horizontal velocity. Therefore, $u_x/U$ is driven to zero one particle diameter, $d$, away from the rod surfaces, which is also marked as a dotted line.

In Fig. \ref{fig:ux-x}(a), we compare the velocity profiles in front of the leading rod ($-x_\ell/D$) at various separation distances. The velocity profile for a single rod is also provided as a baseline, denoted as the solid, black line. For a single rod, the velocity decreases while approaching the rod. This grain slowdown indicates a possible pile-up region, which has been observed in other systems \cite{chehata03, gravish14, kobayakawa2018local}. Adding a second rod, there is significant variation in the profile behavior depending on the separation, however, the velocity near the surface at $s/D=3$ and $s/D=5$ is lower than a single rod at higher separations. The introduction of the second rod increases the pile up region in front of the leading rod, which helps to understand the force behavior observed in Fig.~\ref{fig:all}(a-c).

Next, we compare the velocity profiles behind the leading rod ($+x_\ell/D$) at various separation distances. Here, we find a distinct correlation between the velocity profiles and the separation distance. At low separations, $s/D=3$, the velocity profile is significantly lower than what we find in the single rod. Interference with the following rod causes the velocity behind the leading rod to increase at a slower rate. However, increasing $s/D$, the velocity profiles begin to approach the same form as the profile of a single rod. Once the separation reaches $s/D=18$ for two tandem rods, the flow is nearly identical to that of a single rod. This correlates well with the experimental force data shown in Fig. \ref{fig:all}(b) for $J=1\times10^{-4}$, such that the leading rod becomes independent of the following rod at $s/D>10$.

In Fig. \ref{fig:ux-x}(b), we compare the velocity profiles in front of the following rod ($-x_f/D$) at various separation distances. We find a similar correlation as the discussion for Fig.~\ref{fig:ux-x}(a) in that the leading rod significantly alters the flow field around the following rod at small separation distances. However, at $s/D=20$, the velocity profile in front of the following rod is nearly identical to that of a single rod, indicating that the following rod is independent from the leading rod.

In Fig. \ref{fig:ux-x}(b), we compare the velocity profiles behind the following rod ($+x_f/D$) at various separation distances. We note that the behavior of the velocity profiles seem independent from the separation distance. This leads us to believe that the pressure behind the following rod has little influence on the forces of the leading rod. 

\section{Discussion}
The drag acting on a rod moving relative to a fluid broadly consists of pressure drag and skin drag. The former arises due to the size and shape of the rod, and the latter due to the friction between the rod and the medium~\cite{tritton59}. In a rod moving at similar speeds in similar mediums, it has been found that the pressure drag is significantly larger than the skin drag~\cite{guillard2014lift}. Thus, we rationalize the observed drag acting on the rods 
based on the observations of the velocity field around the rods. \textcolor{black}{We initially noted the role of inter-particle friction as a possible driver for the observed phenomenon. Here, we consider that the inter-particle friction contributes to the stress field gradients around the rods in an imposed flow.} The stress field in the medium due to the flow can be obtained by taking the convective derivative of the local momentum inside the medium obtained by multiplying the medium density and the velocity field. Further, since the flow in the frame of reference of the rods is time-independent, and assuming that the medium density is approximately constant, one can obtain the stress component variation around the intruder from the appropriate spatial derivative. Thus, the stress $\sigma_{xx}$ on the medium due the presence of the rods in the flow direction $x$ and along the axis of symmetry, is obtained from the $x$-derivative of the velocity component $u_x$, i.e. $\sigma_{xx} = -\rho du_x/dx$. Hence, we expect $\sigma_{xx}$ to increase as $u_x$ decreases as the flow approaches the rod as shown in Fig.~\ref{fig:ux-x}(a). Whereas, the stress can be expected to be lower behind the rod as the medium moves around and away from it and is thus extensional. (While we are not aware of calculations of stress on a rod moving vertically through granular medium, a study on rods moving horizontally across a medium show stress distributions consistent with this argument~\cite{guillard2014lift}.) 

\begin{figure*}
    \begin{center}
	    \includegraphics[width=1.5\columnwidth]{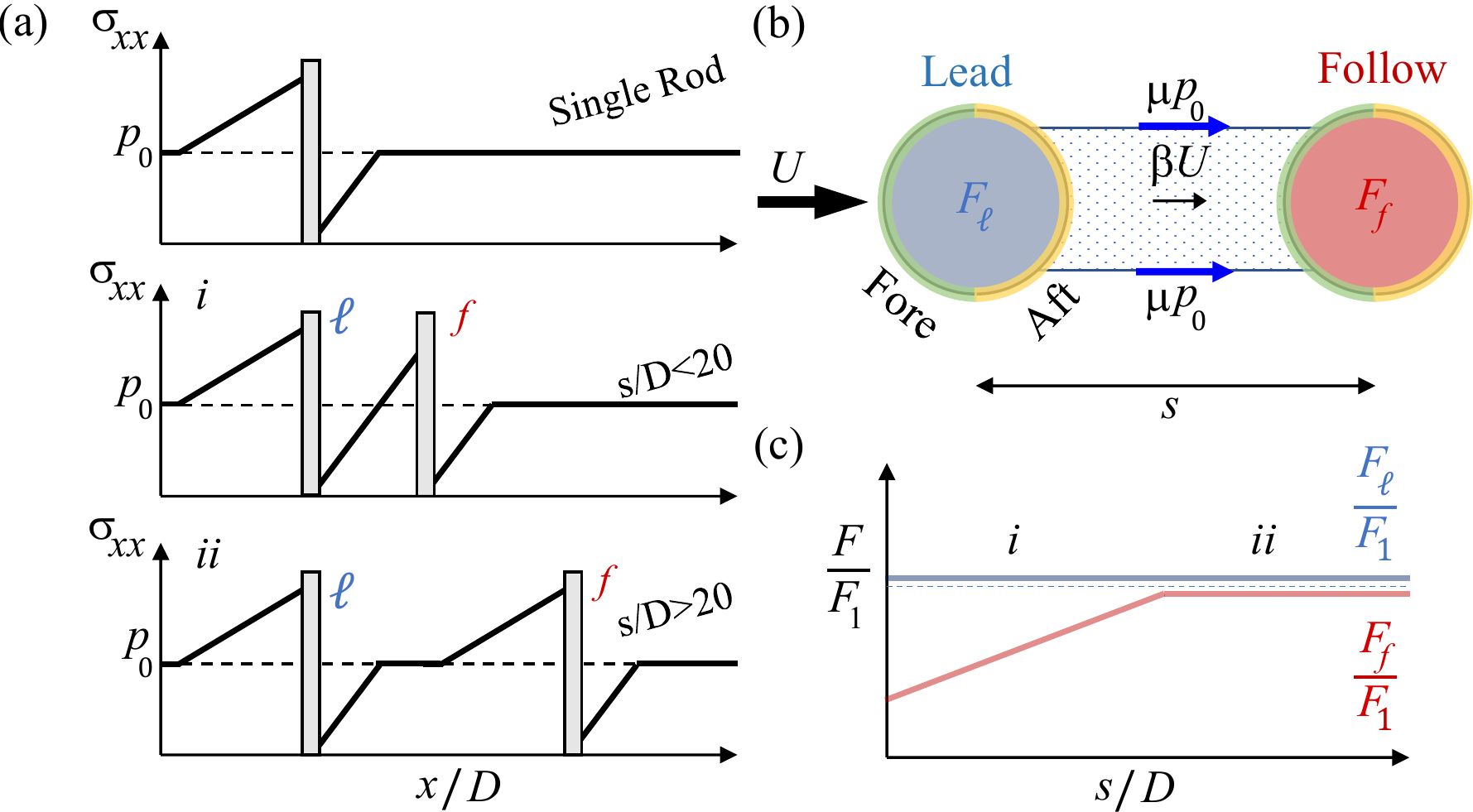}
	\end{center}
	\caption{(a) Sketch of the stress $\sigma_{xx}$ along the axis of symmetry along the flow direction for a single rod, and for rods separated at distances when a stagnant region is (i), and is not observed (ii). The pressure $p_0$ related to the overburden pressure (dashed line) is also provided for reference. The gray rectangles represent the position of rods.  (b) Schematic of the granular flow past two tandem rods with a stagnant zone in between them. The fore-face and the aft-face of the rod surface is highlighted in green and yellow, respectively. $\beta=0$ represents stagnation in which the region between the rods is essentially stationary due to shielding by the two rods.  (c) The drag on the leading rod is not influenced by the flow in the aft-region beyond, but the drag on the following rod increases as width of the stagnant region decreases, and $\beta$ increases with increasing $s/D$.   
	    }
    \label{fig:model}
\end{figure*}
Fig.~\ref{fig:model}(a) shows a sketch of the stress profile for a single rod relative to some reference pressure, $p_0=1/2\phi \rho g z$ at depth $z$, as a function of position $x/D$ along the axis of flow symmetry. Here, we have assumed that pressure is isotropic sufficiently far from the influence of the rods, and given by the overburden pressure. $\sigma_{xx}$ increases over some distance in front of the rod above $p_0$, and then from a lower value at the aft-surface as the stress recovers over some distance back to $p_0$. From the flow measurements we expect these variations to be over $10D$, fore and aft of the rods (see Fig.~\ref{fig:ux-x}). Now, if a following rod is present, one can expect the stress aft of the lead rod to build up more rapidly or similarly, depending on $s/D$, as also sketched in Fig.~\ref{fig:model}(a). The fact that the velocity profiles fore of the leading rod and aft of the following rod collapse onto the single rod data implies that $\sigma_{xx}$ is more or less the same irrespective of $s/D$ in those regions.  

To draw more intuition on the effect of the region between the tandem rods, we turn to a crude approximation of the flow around them in Fig.~\ref{fig:model}(b). We assume there is a region in between the two rods that moves with a velocity $\beta U$, where $\beta$ is a parameter that approximates the relative velocity of the region to the imposed velocity.  When the medium in this region moves essentially with the same speed as the rods, $\beta \approx 0$. Whereas $\beta \approx 1$, as the medium returns to moving with speed $U$ after the lead rod for sufficiently large enough $s$~\footnote{From Fig.~\ref{fig:ux-x}, one can note that $\beta \approx 0.2$ for $s/D = 3$, and $\beta \approx 0.4$ for $s/D = 5$, whereas for $s/D = 11$ and 20, we have $\beta \approx 1$. Further from Fig.~\ref{fig:ux-y}(a), one can note that the width of the slow flowing region is $\approx D$ for $s/D = $ and 5, but is essentially zero for $s/D = 11$ and $20$.}.
We surmise that the width of the stagnant region $w_s$ is a fraction of the diameter of the rods which decreases starting from $w_s/D = 1$ to $w_s/D = 0$, as $s/D$ increases. For efficacy, we further simply assume that $\beta = 0$ inside the stagnant region, and the flow speed is $U$ outside. Thus, while the entire fore-face of the leading rod is exposed to the medium flowing with speed $U$, only a fraction of the fore-face of the following rod, given by $1 - w_s/D$, is exposed to the flow. It should be noted that there is little relative flow between the stagnant region and the following rod, and thus may not be expected to contribute the form drag. However, because of the shear applied by the flowing region on the stagnant region given by the coefficient of friction $\mu$ of the granular medium, the shielded zone also can be expected to contribute the drag which increases initially with $s$ till the width $w_s$ decreases to zero.     

Taken together, these arguments imply that drag $F_\ell$ acting on the leading rod is essentially constant with $s/D$ (see Fig.~\ref{fig:model}(c)). Whereas, the drag $F_f$ acting on the following rod increases slowly from the near zero value it encounters when it is nearly fully shield behind the leading rod. $F_f$ can be then expected to approach the same value as $F_\ell$ for sufficiently large $s/D$. These arguments capture to first approximation the main features of the observed drag $F_\ell$ and $F_f$ in Fig.~\ref{fig:all} in granular beds with the various interstitial fluids. A more detailed model and quantitative analysis is needed to capture the subtle increases of $F_\ell$ above even the single rod value at short distances, and the detailed $s/D$-dependence. \textcolor{black}{Future studies with a fluidized bed may elucidate the role of inter-particle friction on the medium's stress field and force distribution on the rods.} 

\section{Conclusions}
By performing experiments with rods in tandem, we demonstrate that the drag acting on the rods is nonadditive for sufficiently small separation distances in dry granular beds, as well as in beds fully immersed in a viscous Newtonian liquid. We find that the total drag acting on the rods is nearly half the drag at large separations, and is nearly equal to the drag of a single rod in all the various granular and fluid combinations investigated. By measuring the drag experienced by each rod, we show that the lead rod continues to experience more or less the same drag, irrespective of the presence or location of the following rod. Whereas, the following rod is found to be effectively shielded by the leading load, and experiences little drag at small separation distances. As separation distance increases, the drag acting on the following rod increases and is found to approach the value for an independent rod. 

The observed drag in the granular beds was further compared with measurements with a purely viscous Newtonian fluid (corn syrup) to understand the effect of the nature of the medium on the observed drag. By matching the effective flow Reynolds numbers, we find that the total or net drag acting on the rods are nearly identical across the different mediums with separation distance. However, the drag acting on the leading and following rod are nearly identical in the case of the Newtonian fluid since our experiments focus on the low Reynolds regime. Thus, we conclude that the drag observed in granular mediums is different from that in a viscous fluid in spite of the fact that the total drag as a function of separation distance behaves somewhat similarly. 

By visualizing the flow around the rods in the granular medium, as well as in the Newtonian fluid, we further find both qualitative and quantitative differences consistent with the observed drag measurements. The perturbation of flow due to the rods is observed to be more narrowly confined around the intruders compared with the viscous fluids, i.e. the flow decays more rapidly to the far field limit in the granular case. We find that while the flow in front of the leading rod, and behind the following rod are more or less unchanged with separation distance, a significant stagnation of the flow occurs in the region between the rods with decreasing separation. This shielding of the flow by the leading rod results in the lower drag of the following rod. Further modeling work is required to quantitatively capture the observed drag as a function of separation distance. 

More broadly, the fact that the measured drag acting on these tandem rods does not sum up linearly at short distances means that, linear superposition - which is at the heart of resistive force theory~\cite{hancock81,li13} used to calculate drag acting on extended or multi-component objects such as limbs of burrowing animals - is not strictly observed in granular mediums, just as previously noted in viscous Newtonian fluids. The further observation that the burden of drag is vastly differently distributed between the leading and following rods is further reason for caution. In the configuration of two parallel rods intruding into a granular bed, we note a similar behavior of nonadditive forces occurs in a small range of separation distances \cite{Pravin2021}. Thus, in spite of recent success of RFT in capturing drag in simple shaped objects moving in dry~\cite{gravish14} and immersed granular medium~\cite{allen19,kudrolli2019burrowing,pal22}, care should be exercised in applying RFT to multi-limb entities moving through granular mediums.

\section{Acknowledgments}
We thank Benjamin Allen for discussions and performing preliminary experiments.
This work was supported by National Science Foundation CBET Grant under Grant No. CBET-1805398. 


\end{document}